\documentclass[12pt,preprint]{article}
\usepackage{amssymb}
\usepackage{amsmath}
\usepackage{graphics}
\usepackage{epsfig}

\newcommand{\eqb}{\begin{equation}}
\newcommand{\eqe}{\end{equation}}
\newcommand{\dmb}{\begin{displaymath}}
\newcommand{\dme}{\end{displaymath}}

\newcommand{\eab}{\begin{eqnarray}}
\newcommand{\eae}{\end{eqnarray}}

\newcommand{\e}{\mbox{e}}
\newcommand{\be}{\begin{equation}}
\newcommand{\ee}{\end{equation}}

\input{tcilatex}

\begin{document}

\begin{titlepage}
\begin{flushright} 
\end{flushright}
\vspace{0.6cm}

\begin{center}
\Large{Onset of magnetic monopole-antimonopole condensation}
\vspace{1.5cm}

\large{Ralf Hofmann}

\end{center}
\vspace{1.5cm} 

\begin{center}
{\em Institut f\"ur Theoretische Physik\\ 
Universit\"at Heidelberg\\ 
Philosohenweg 16\\ 
69120 Heidelberg, Germany}
\end{center}
\vspace{1.5cm}

\begin{abstract}

We determine the critical strength of the effective electric coupling for the onset of Bose 
condensation of stable magnetic monopoles and antimonopoles in 
SU(2) Yang-Mills thermodynamics. Two scenarios are considered: infinitely fast and infinitely slow downward 
approach of the critical temperature. Our results support 
the claim that the first lepton family and the weak interactions emerge from pure 
SU(2) gauge dynamics of scale $\sim 0.5\,$MeV.  

\end{abstract} 

\end{titlepage}

\noindent{\sl Introduction.} A genuine understanding of fun\-damental physics requires well-con\-trolled con\-ditions. As for 
four-dimensional Quantum Yang-Mills theory unadulterated thermodynamics seems to be such a 
setting \cite{Hofmann2005,Hofmann2007}. Because the term `understanding' not only refers to the ability to derive 
quantitative results but also to one's position to actually interpret them in 
deterministic terms a connection between effective (results) and 
fundamental (interpretation) fluctuations in the according formulations of the same 
partition function needs to be made. This connection also serves as the pointer to 
how a controlled deformation of the original thermodynamical setting can be carried out. 

In this note we are concerned with the transition between deconfining and preconfining SU(2) 
Yang-Mills thermodynamics. In approaching this transition from above in a specified way, 
we ask the question what 
the critical value for the effective, electric coupling $e$ is at 
which the thermal ground state of the deconfining phase drastically 
rearranges such as to attribute mass to a formerly massless gauge mode 
(condensation of magnetic (anti)monopoles). Technically speaking, the answer to this question is 
obtained in a surprisingly simple way if the above-mentioned connection between effective and 
fundamental fluctuations is made. Starting from a useful a priori estimate of the deconfining 
thermal ground state the computation of thermodynamical quantities is organized into a rapidly 
converging loop expansion carried by effective gauge-field fluctuations of 
trivial topology. At high temperatures this expansion indicates 
the presence of stable and isolated (anti)monopoles of a given number density, 
and at temperatures not far above the critical temperature $T_c$ the collective dynamics 
of these defects, essentially exhausted by two-loop corrections, induce, depending on momentum, 
screening or antiscreening of the tree-level massless, effective gauge mode \cite{HHR2004,SHG2006,LH2008,LKGH2008}.     

\noindent{\sl Some remarks on deconfining SU(2) Yang-Mills thermodynamics.} 
On the one-loop level the Yang-Mills system is approximated by a gas 
of noninteracting thermal quasiparticles fluctuating above 
a thermal ground state. The latter is decribed by an effective inert, adjoint scalar field and a 
pure-gauge configuration, and the mass spectrum of thermal quasiparticles is 
made explicit in admissible unitary gauge 
\cite{Hofmann2005,Hofmann2007}. On the fundamental level, 
this situation locally is induced by 
interacting (anti)calorons of topological charge-modulus unity 
whose small holonomy \cite{Nahm,LeeLu,KraanVanBaal}, temporarily created by the absorption 
of soft and fundamental propagating gauge fields, is insufficient for the permanent 
release of their magnetic monopole-antimonopole constituents 
\cite{Diakonov2004}. This approximation captures the total pressure up to an error smaller 
than one percent. Effective radiative corrections describe a 
departure from this situation in so far as the thermal ground 
state is contributed to by domainized configurations of the adjoint scalar field 
with the vertices of sufficiently many domain boundaries \cite{Kibble} 
representing magnetic (anti)monopoles. At high temperature, an average over such configurations is performed 
implicitly by a particular two-loop diagram for the pressure \cite{HHR2004,SHG2006} whose 
ratio to the one-loop approximation with increasing 
temperature rapidly approaches a small, negative constant. This constant represents the existence of an 
average density of highly nonrelativistic and screened 
magnetic (anti)monopoles which are released by the rare and irreversible 
dissociation of (anti)calorons \cite{LKGH2008,Diakonov2004}. The irreversibility of (anti)caloron 
dissociation together with the fact that overall magnetic charge is nil 
due to pairwise monopole-antimonopole creation implies that the chemical potential associated with monopole-antimonopole 
pairs vanishes in the infinite-volume limit.   

As temperature decreases, stable (anti)monopoles behave like comoving raisins, immersed in an expanding, infinitely extended 
dough, with their average distance set by the inverse 
temperature. In the hypothetic limit of isolation, see \cite{LKGH2008}, 
the liability of a caloron or an anticaloron to dissociate into a pair of an 
isolated monopole of mass $m_M$ and its antimonopole of mass $m_A$ is 
determined solely by its holonomy \cite{LeeLu,KraanVanBaal,Diakonov2004}. For the realistic 
case of densely packed (anti)calorons the description of a single caloron 
by semiclassical methods fails \cite{LKGH2008}.  An {\sl average} over monopole-antimonopole creation 
processes, however, leads to a number density $n_{M+A}$ of pairs which is determined by 
the following holonomy-independent sum of masses \cite{LeeLu,KraanVanBaal}
\eqb
\label{monopmass}
m_{M+A}=m_M+m_A=\frac{8\pi^2 T}{e(\lambda)}+\mbox{non-BPS}\,,
\eqe
where $T$ and $\lambda\equiv\frac{2\pi T}{\Lambda}$ are the dimensionful and the dimensionless versions of 
the temperature, respectively, and $\Lambda$ denotes the Yang-Mills scale. The $\lambda$ dependence of 
$e$ is a consequence of the renormalization-group invariance of the a priori estimate of the Yang-Mills 
partition function under the spatial coarse-graining applied to derive the effective 
theory \cite{Hofmann2005,Hofmann2007}. This running of $e$ with temperature describes the screening effects due 
to instable magnetic dipoles arising from (anti)calorons whose holonomy is 
only mildly deformed away from trivial. Notice that $e$ approaches a 
plateau $e\equiv\sqrt{8}\pi$ very rapidly with increasing $\lambda>\lambda_c=13.87$. For low temperatures, $\lambda_c\le\lambda\le 15.0$, 
we have\footnote{The author would like to thank Markus Schwarz for performing this fit.}
\eqb
\label{lambdapole}
e(\lambda)=-4.59\,\log(\lambda-\lambda_c)+18.42\,.
\eqe
In Eq.\,(\ref{monopmass}) the term `non-BPS' refers to the effects which are induced by the presence of all other 
isolated, stable and screened magnetic monopole-antimonopole pairs which, in contrast to the situation of screening by instable dipoles, 
introduce a mass scale into the decay properties 
of the magnetic potential of a given stable (anti)monopole. 
The correction to the BPS mass should be comparable to the dual gauge mode's magnetic screening mass $m_m$ which due to weak coupling is 
calcu\-lable in per\-tur\-bation 
theory\footnote{This yields an order-of-magnitude 
result which matches well with the exact result extracted from a two-loop correction of the pressure at high 
temperature \cite{LKGH2008}}. To lowest order in the magnetic coupling $g=4\pi/e$ we have 
$m_m=\frac{4\pi T}{\sqrt{3}\,e}$ \cite{Andersen}. Compared to the BPS term in Eq.\,(\ref{monopmass}) this is a 
correction of less than 10\,\%.

\noindent{\sl Condensation of monopole-antimonopole pairs.} Based on the discussion presented by Huang \cite{Huang} of thermalized, 
noninteracting Bose particles with
 mass $m$ a relation was formulated in \cite{Nieto} between the total number density $n$ and 
the density $n_0$ of particles residing in the condensate. For statistical weight unity (only one species 
of monopole-antimonopole pairs occurs in an SU(2) Yang-Mills theory) 
one has
\eab
\label{condcond}
n_0&=&n-n_c\nonumber\\ 
&\equiv&n-\frac{T^3}{2\pi^2}\,\mu^2\,\sum_{l=1}^{\infty} \e^{l\mu}\,\frac{K_2(l\mu)}{l}\,,
\eae
where $\mu\equiv m/T$, and $K_2(x)$ is the modified Bessel function of the second kind. 
At the onset of Bose condensation, where $n_0$ is yet zero, 
the total number density $n$ is given by the number density $n_{\tiny\mbox{fr}}$ 
of freely fluctuating particles  
\eqb
\label{numdens}
n=n_{\tiny{\mbox{fr}}}\equiv\frac{T^3}{2\pi^2}\,\int_0^\infty dx\,\frac{x^2}{\e^{\sqrt{x^2+\mu^2}}-1}\,.
\eqe
So in the fully thermalized system, which takes place if $T$ is slowly lowered towards $T_{\tiny\mbox{c}}$, 
we have at the onset of Bose condensation
\eqb
\label{num}
\int_0^\infty dx\,\frac{x^2}{\e^{\sqrt{x^2+\mu_{\tiny\mbox{c}}^2}}-1}=n_c=
\mu^2_{\tiny\mbox{c}}\,\sum_{k=1}^{\infty} \e^{k\mu_{\tiny\mbox{c}}}\,\frac{K_2(k\mu_{\tiny\mbox{c}})}{k}\,.
\eqe
Eq.\,(\ref{num}) determines the 
critical {\sl ratio} $\mu_{\tiny\mbox{c}}$ between mass and temperature 
at which Bose condensation starts to 
occur. For conventional condensed-matter systems 
the mass $m$ of a given species of bosonic particles is a predetermined 
quantity which does not depend on 
temperature in the absence of interactions. 
The solution $\mu_{\tiny\mbox{c}}$ to Eq.\,(\ref{num}) thus determines the critical temperature $T_{\tiny\mbox{c}}$
for Bose condensation as $T_{\tiny\mbox{c}}=\frac{m}{\mu_{\tiny\mbox{c}}}$. 
In deconfining Yang-Mills thermodynamics, however, a pair of an isolated and screened magnetic monopole and its antimonopole 
owes its very existence to the presence of a heat bath of 
given temperature. The according relation between mass and temperature, see Eq.\,(\ref{monopmass}), 
together with $\mu_{\tiny\mbox{c}}$ determines the critical temperature 
for condensation to be the solution to the following equation 
\eqb
\label{critT}
 T_{\tiny\mbox{c}}=\frac{m(T_{\tiny\mbox{c}})}{\mu_{\tiny\mbox{c}}}\,.
\eqe
Notice that in the limit $\mu_{\tiny\mbox{c}}\to 0$ Eq.\,(\ref{num}) yields the identity $2\,\zeta(3)=2\,\zeta(3)$ where 
$\zeta(z)$ is Riemann's zeta function. Since the left-hand side of Eq.\,(\ref{num}) is 
monotonic decreasing and the right-hand side is 
monotonic increasing in $\mu_{\tiny\mbox{c}}$ it follows that $\mu_{\tiny\mbox{c}}=0$ is the only solution. 
Thus in deconfining SU(2) Yang-Mills thermodynamics (adiabatically slow approach of $T_c$ from above) 
only massless monopoles and antimonopoles condense into a new ground state 
at the critical temperature $T_c$ corresponding to the logarithmic pole in $e$ described by Eq.\,(\ref{lambdapole}).  

Alternatively, one may ask the question of what happens in the limit where $T_c$ is rapidly approached from above. As we will see, such an 
adiabatic (sudden) approximation fully takes into account the static screening effects imposed by the system at high temperatures 
but neglects the influence of propagating dual gauge modes not too far above $T_c$. Since the pole of the coupling $e(\lambda)$, 
see Eq.\,(\ref{lambdapole}), is logarithmic (reflecting the fact that the Yang-Mills scale 
$\Lambda$ nonperturbatively interferes with the dynamics of fundamental propagating gauge 
modes only shortly above $\lambda_{\tiny\mbox{c}}$ \cite{GH2007}) we may consider the limit of large temperatures 
for the dependence on temperature of the density $n_{M+A,\tiny\mbox{as}}$ of {\sl interacting} but statically equilibrated 
monopole-antimonopole pairs. This situation is relevant for particle collisions at sufficiently high 
center-of-mass energy where locally a hot spot of deconfining phase is generated whose temperature quickly drops due to cooling and shrinking by 
evaporation \cite{GHM2008}.    

Recall that $n_{M+A,\tiny\mbox{as}}$ is extracted from a 
particular two-loop correction to the quasiparticle pressure as calculated 
in the effective theory \cite{LKGH2008}. One has
\eqb
\label{asymptdens}
n_{M+A,\tiny\mbox{as}}= (21.691)^{-3}\,T^3\sim 9.8\times 10^{-5}\,T^3\,.
\eqe
At high temperatures isolated, stable and screened (anti)monopoles are nonrelativistic \cite{LKGH2008}. 
If temperature is lowered towards $T_{\tiny\mbox{c}}$ in a sufficiently rapid 
way then the generation of almost massless, stable monopoles and antimonopoles by 
strong screening occurs quickly enough to not affect their 
nonrelativistic nature endowed by high-temperature physics\footnote{To catch up in velocity (anti)monopoles must 
interact via the exchange of dual gauge modes that 
are close to their mass shell and therefore propagate at a speed close to the velocity of light. 
In contrast to the portion of magnetic screening induced by an 
increased activity of instable monopole-antimonopole pairs and described by the effective-theory a priori estimate for the 
thermal ground state this part of the thermalization of (anti)monopoles -- a loop correction in the effective theory -- 
thus requires a finite amount of time.}. The condensation condition (\ref{num}) thus modifies as
\eqb
\label{condconmod}
\frac{n_{M+A,\tiny\mbox{as}}}{T^3}=(21.691)^{-3}=\zeta(3/2)\,\left(\frac{\mu_{\tiny\mbox{c},M+A}}{2\pi}\right)^{3/2}\,.
\eqe
where the expression to the far right is obtained by considering $\mu_{\tiny\mbox{c},M+A}\gg 1$ of $n_c$ or, equivalently, of 
$1/2\pi^2$ times the right-hand side of (\ref{num}). To summarize, 
Eq.\,(\ref{condconmod}) determines $\mu_{\tiny\mbox{c},M+A}$ in a situation where 
highly nonrelativistic, stable and isolated (anti)monopoles are adiabatically 
fast deprived of their mass by cooling (enhanced instantaneous screening by instable dipoles) 
so that no time is available for them to start moving.  

The solution to Eq.\,(\ref{condconmod}) is $\mu_{\tiny\mbox{c},M+A}=7.04\times 10^{-3}$. 
Note the amusing fact that the nonrelativistic nature of 
monopole-antimonopole pairs is assured by the large-mass limit of $n_c$ while the solution to Eq.\,(\ref{condconmod}) 
actually corresponds to a small 
mass on the scale of temperature. The resolution of this apparent puzzle is grounded in the fact that 
the sudden approximation employed does not admit a thermodynamical interpretation: the expression for 
$n_{M+A,\tiny\mbox{as}}$ at $T\gg T_c$ is analytically continued down to $T_c$.     

Using
\eqb
\label{mucorr}
\mu_{M+A}=\left(8\pi^2+\frac{4\pi}{\sqrt{3}}\right)\,e^{-1}\,,
\eqe
compare with Eq.\,(\ref{monopmass}) and paragraph below Eq.\,(\ref{lambdapole}), 
we obtain 
\eqb 
\label{crite}
e_{\tiny\mbox{c}}=1.225\times 10^4\,.
\eqe
The result in Eq.\,(\ref{crite}) acts as a lower bound 
for the values of $e_{\tiny\mbox{c}}$ occurring for finite-velocity 
approaches $\lambda\searrow\lambda_c$. That is, for this nonadiabatic 
situation $\mu_{\tiny\mbox{c},M+A}$ must take values inbetween the extremes obtained 
at zero and infinite velocity:
\eqb
\label{estmation}
0\le\mu_{\tiny\mbox{c},
+A}\le 7.04\times 10^{-3}\ \ \ \ \ \ \mbox{or}\ \ \  \ \ 
\infty\ge e_{\tiny\mbox{c}}\ge 1.225\times 10^4\,.
\eqe    

\noindent{\sl Mass of charged vector bosons in the Standard Model.} The ratio of charged-vector-boson-mass $m_W$ to 
electron mass $m_e$, as experimentally measured, is given as
\eqb
\label{ratiomasses}
\frac{m_W}{m_e}=1.6\times 10^5\,.
\eqe
If we postulate that a pure SU(2) Yang-Mills theory of Yang-Mills scale 
$\Lambda_e\sim m_e$ is responsible for the emergence of the electron and its neutrino in its 
confining phase \cite{Hof07,MH081,MH082,GHM2008} and for the mediation of the weak force by its decoupling, dynamically 
massive gauge bosons \cite{Hofmann2005} ($W^\pm$ at the deconfining-preconfining transition and $Z^0$ at the 
preconfining-confining transition) then the value of $e$ as 
calculated from the ratio in Eq.\,(\ref{ratiomasses}) should be contained in the 
range specified by (\ref{estmation}). 

Let us check whether this indeed is the case. Since in such a theory 
one would have 
\eqb
\label{mWsu2}
m_W=2e_{\tiny\mbox{c}}\,|\phi|(T_{\tiny\mbox{c}})=
2e_{\tiny\mbox{c}}\,\Lambda_e\,\lambda^{-1/2}_{\tiny\mbox{c}}=
e_{\tiny\mbox{c}}\,\Lambda_e\,\sqrt{\frac{4}{13.87}}
\eqe 
and since $\Lambda_e\sim m_e$ the value of $e_{\tiny\mbox{c}}$ should relate to the experimentally determined ratio in Eq.\,(\ref{ratiomasses}) as 
follows
\eqb
\label{eexp}
e_{\tiny\mbox{c}}\sim \frac{m_W}{m_e}\,\sqrt{\frac{13.87}{4}}=2.98\times 10^{5}\,.
\eqe
Obviously, this value for $e_{\tiny\mbox{c}}$ lies in the range given by (\ref{estmation}). 

\noindent{\sl Summary.} In this note we have considered two scenarios for the onset of 
magnetic monopole-antimonopole condensation at the deconfining-preconfining 
transition in SU(2) Yang-Mills thermodynamics: Infinitely slow and infinitely fast downward 
approach of $T_c$. In the former situation, we have shown that pairs of stable monopoles and 
antimonopoles do only condense when they are massless, that is, at the pole position for the effective electric coupling $e$. 
This is a consequence of the fact that due to the irreversibility of the (anti)monopole creation 
process (dissociation of large-holonomy (anti)calorons) and due to overall charge neutrality (pairwise creation in an infinite spatial volume) 
the chemical potential associated with pairs is nil. Concerning the case of infinitely fast approach of $T_c$, 
we obtain a lower bound on the value $e_{\tiny\mbox{c}}$ 
of the critical coupling. Our results are compatible with the claim that a pure SU(2) Yang-Mills theory 
of scale $\Lambda_e\sim m_e\sim 0.5$\,MeV is responsible for the emergence of the first lepton family 
and the weak interactions of the Standard Model of Particle Physics. 
Due to the experimental fact of a universal electric coupling of the photon -- likely 
to emerge from an SU(2) Yang-Mills theory of scale $\Lambda_{\tiny\mbox{CMB}}\sim 10^{-4}\,$eV \cite{SHG2006,LH2008,Hofmann2009} -- 
to all charged leptons it is clear that Nature's SU(2) Yang-Mills theories of the same electric-magnetic parity mix maximally. 

\section*{Acknowledgments}
We would like to acknowledge useful conversations with Markus Schwarz.


\begin{thebibliography}{99}

\bibitem{Hofmann2005} R. Hofmann, Int. J. Mod. Phys. A \textbf{20}, 4123
(2005), Erratum-ibid. A \textbf{21}, 6515 (2006). (arXiv:hep-th/0504064)

\bibitem{Hofmann2007}
R. Hofmann, (arXiv:0710.0962 [hep-th]). 

\bibitem{HHR2004}
U. Herbst, J. Rohrer, and R. Hofmann, Acta Phys. Polon. B{\bf 36}, 881 (2005). (arXiv:hep-th/0410187)

\bibitem{SHG2006}
M. Schwarz, R. Hofmann, and F. Giacosa, Int. J. Mod. Phys. A \textbf{22}, 1213 (2007). 
(arXiv:hep-th/0603078)

\bibitem{LH2008}
J. Ludescher and R. Hofmann, Ann. Phys. (Berlin) {\bf 18}, 271 (2009).(arXiv:0806.0972 [hep-th])

\bibitem{LKGH2008}
J. Ludescher, J. Keller, F. Giacosa, and R. Hofmann, arXiv:0812.1858 [hep-th].

\bibitem{Nahm}
W. Nahm, Phys. Lett. B {\bf 90}, 413 (1980).\\
W. Nahm, {\sl The ADHM construction for instantons, self-dual monopoles and calorons},
Lect. Notes in Physics, Vol. 201, eds. G. Denaro, Springer Heidelberg, 189 (1984).

\bibitem{LeeLu}
K.-M. Lee and C.-H. Lu, Phys. Rev. D {\bf 58}, 025011 (1998). (arXiv:hep-th/9802108)

\bibitem{KraanVanBaal}
T. C. Kraan and P. van Baal, Nucl. Phys. B {\bf 533}, 627 (1998). (arXiv:hep-th/9805168)\\
T. C. Kraan and P. van Baal, Phys. Lett. B {\bf 435}, 389 (1998). (arXiv:hep-th/9806034)

\bibitem{Diakonov2004}
D. Diakonov, N. Gromov, V. Petrov, and 
S. Slizovskiy, Phys. Rev. D {\bf 70}, 036003 (2004). (arXiv:hep-th/0404042)

\bibitem{Kibble}
T.W.B. Kibble, J. Phys. A {\bf 9}, 1387 (1976). 

\bibitem{Andersen}
J. O. Andersen, Z. Phys. C \textbf{75}, 147 (1997). (arXiv:hep-ph/9606357)

\bibitem{Huang}
K. Huang, {\sl Statistical Mechanics} (John Wiley \& Sons, Inc. New York, 1963), Sec. 9.6 and 12.3.  

\bibitem{Nieto}
M. M. Nieto, J. Math. Phys. \textbf{11}, 1346 (1970).\

\bibitem{GHM2008}
J. Moosmann and R. Hofmann, arXiv:0908.1502 [hep-ph]. 

\bibitem{GH2007}
F. Giacosa and R. Hofmann, Phys. Rev. D {\bf 77}, 065022 (2008). (arXiv:0704.2526 [hep-th])

\bibitem{Hof07}
R. Hofmann, Mod. Phys. Lett. A {\bf 22}, 2657 (2007). (arXiv:hep-th/0702027)

\bibitem{MH081}
J. Moosmann and R. Hofmann, arXiv:0804.3527 [hep-th]. 

\bibitem{MH082}
J. Moosmann and R. Hofmann, arXiv:0807.3266 [hep-th].



\bibitem{Hofmann2009}
R. Hofmann, Ann. Phys. (Berlin) {\bf 18}, 634 (2009). (arXiv:0902.2700 [hep-th])

\end{thebibliography}
\end{document}